\documentstyle[12pt]{article}
\newcommand{\tilt}{\tilde{t}_{\alpha}}
\newcommand{\tilj}{\tilde{J}_{\alpha}}
\newcommand{\opn}{\hat{n}}
\newcommand{\vk}{{\bf k}}
\newcommand{\vq}{{\bf q}}
\begin{document}
\baselineskip=1.6 cm

\begin{center}
{\huge Superconductivity in a Quasi One Dimensional Spin Liquid}
\end{center}

\baselineskip=0.6 cm
\vskip 1.5 cm

\begin{center}
M. Sigrist$ ^{a,b)} $, T.M. Rice$ ^{b)} $ and F.C. Zhang$ ^{c)}$
\end{center}

\vskip 0.4 cm

\begin{center}
a) Paul Scherrer Institut, 5232 Villigen PSI, Switzerland \\
b) Theoretische Physik, ETH-H\"onggerberg, 8093 Z\"urich, Switzerland
\\
c) Department of Physics, University of Cincinnati, Cincinnati,Ohio
45221
\end{center}

\vskip 1.5 cm

\noindent
{\bf Abstract:} The single rung t-J ladder is analyzed in a
mean field theory using Gutzwiller renormalization
of the matrix elements to account for strong correlation. The spin
liquid (RVB) state at half-filling evolves into a superconducting
state upon doping. The order parameter has a modified d-wave
character. A lattice of weakly coupled ladders should show a
superconducting phase transition.

\vskip 0.5 cm


\vskip 0.8 cm

\noindent
There is a striking difference between the properties of a chain and
a ladder (double chain) antiferromagnetic (AF) s=1/2 Heisenberg model.
Whereas the chain has power law decay of the AF-correlations, the ladder has a
purely exponential decay and a finite energy gap in the spin
excitation spectrum, i.e. a spin gap (see for example Ref.1). If, as
is the case for other
spin gap systems, the spin gap persists to finite doping, then the
possibilities for superconducting fluctuations are greatly enhanced in
ladder systems [2]. Recently we pointed out that the compound $
{\rm Sr}_2 {\rm Cu}_4 {\rm O}_6 $ offers the possibility of realizing
a lattice of weakly coupled ladders [3,4]. This compound
is a member of the homologous series $ {\rm Sr}_{n-1} {\rm Cu}_{n+1}
{\rm O}_{2n} $, which differ from known high-$ {\rm T}_c $ cuprates
through the presence of a parallel array of line defects in the $ {\rm
CuO}_2 $-planes. In this letter we examine the properties of undoped
and lightly doped ladders described by a t-J model within a
mean field approximation which uses a Gutzwiller renormalization
factor to approximate the local constraint. Superconducting
correlations of a modified d-wave symmetry are predicted. The system of
weakly coupled chains offers an interesting example of a short range
resonance valence bond (RVB) state in a system intermediate between 1
and 2 dimensions [5].

The basic model we apply is the t-J model on the ladder
with the Hamiltonian

\begin{equation} \begin{array}{ll}
{\cal H} = & \displaystyle -\sum_{i,{\rm s}} [ t_x \sum_{a=1,2}
c^{\dag}_{i+1,a,{\rm
s}} c_{i,a,{\rm s}} + t_y c^{\dag}_{i,1,{\rm s}} c_{i,2,{\rm s}} +h.c.
] \\ & \\
& \displaystyle + \sum_i [J_x \sum_{a=1,2} {\bf S}_{i+1,a} \cdot {\bf
S}_{i,a} +
J_y {\bf S}_{i,1} \cdot {\bf S}_{i,2} ] \\ & \\
\end{array} \end{equation}

\noindent
where $ i $ runs over all rungs and $ a $ over the
two legs, 1 and 2. The constraint $ \sum_{\rm s} c^{\dag}_{i,a,{\rm
s}} c_{i,a,{\rm s}} \leq 1 $, projecting out all the
doubly occupied states on each site, is implied.
We examine this model within the mean field theory used
by Zhang et al. for the 2D square lattice [6].
The constraint is taken into account approximately by a Gutzwiller
type renormalization of the coupling constant along the chain,
$ \langle c^{\dag}_{i+1,a,{\rm s}} c_{i,a,{\rm s}} \rangle = g_{tx}
\langle c^{\dag}_{i+1,a,{\rm s}} c_{i,a,{\rm s}} \rangle_0 $, and
$ \langle {\bf S}_{i+1,a} \cdot {\bf S}_{i,a} \rangle = g_{Jx}
\langle {\bf S}_{i+1,a} \cdot {\bf S}_{i,a} \rangle_0 $, and analogous
$ g_{ty} $ and $ g_{Jy} $  on the rungs with $ \langle ... \rangle
$ and $ \langle ... \rangle_0  $ the expectation values in the
projected and unprojected state
respectively. These $ g $-factors are determined by the ratio of the
probabilities of the matrix elements in the projected and unprojected
states [6]. The renormalized coupling constants are then defined by $
\tilj = J_{\alpha} g_{J \alpha} $ and $ \tilt
= t g_{t \alpha} $.

Using the fermion representation for the Heisenberg term in
(1), we introduce two types of mean fields,
$ \chi_{x} = \langle c^{\dag}_{i+1,a,{\rm s}} c_{i,a,{\rm s}} \rangle_0 $ and
$ \Delta_{x} = \langle c_{i+1,a,\downarrow} c_{i,a \uparrow}
\rangle_0 $, and $ \chi_{y} = \langle c^{\dag}_{i,1,{\rm s}} c_{i,2,{\rm s}}
\rangle_0 $ and
$ \Delta_{y} = \langle c_{i,1,\downarrow} c_{i,2 \uparrow}
\rangle_0 $, a bond and pairing mean field, respectively.
The t-J model is then reduced to a renormalized Hamiltonian with the
constraint released

\begin{equation} \begin{array}{ll}
{\cal H}_{MF} = & \displaystyle \sum_{\vk,s} \varepsilon_{\vk} c^{\dag}_{\vk,s}
c_{\vk,s} - \sum_{\vk} [\Delta^*_{\vk} c_{\vk \downarrow} c_{-\vk
\uparrow} + h.c.] \\ & \\
& \displaystyle + \frac{3N}{4} \sum_{\alpha=x,y} f_{\alpha}
\tilde{J}_{\alpha} (|\chi_{\alpha}|^2+|\Delta_{\alpha}|^2)
\end{array} \end{equation}

\noindent
where $ f_x=2 $ and $ f_y =1 $ and $ N $ is the number of lattice
sites. The summation over $ {\bf k} $ is restricted to $ -\pi < k_x
\leq \pi $  and $ k_y $ equals to 0 (bonding) or $ \pi $
(antibonding). In (2)

\begin{equation} \begin{array}{ll}
\varepsilon_{\vk} & \displaystyle = - 2 \sum_{\alpha} f_{\alpha}
(\tilde{t}_{\alpha} + \frac{3}{4} \tilde{J}_{\alpha} \chi_{\alpha}) {\rm
cos} k_{\alpha} - \mu \\ & \\
\Delta_{\vk}& \displaystyle = \frac{3}{2} \sum_{\alpha}
f_{\alpha} \tilde{J}_{\alpha} \Delta_{\alpha} {\rm
cos} k_{\alpha}   \\ & \\
\end{array} \end{equation}

\noindent
with $ \mu $ as the effective chemical potential. We calculate the
Gutzwiller renormalization factors by including the correlations of
the probability between the two nearest neighbor sites [7], which improves
the approximation of Ref.6. They are given by

\begin{equation}\begin{array}{ll}
g_{t \alpha} & \displaystyle ={ 2n \delta}/({n(1+\delta) + 4 \chi^2_{\alpha}})
\\ & \\
g_{J \alpha} & \displaystyle= {4 n^2}/({n^2(1+\delta)^2+8
\delta^2 (|\Delta_{\alpha}|^2-|\chi_{\alpha}|^2) + 16
(|\Delta_{\alpha}|^4+|\chi_{\alpha}|^4)}) \\ & \\
\end{array}\end{equation}

\noindent
where $ n $ is the electron density and $ \delta =1-n $ is the hole
doping concentration.

The mean field Hamiltonian has to be solved self-consistently together
with the mean field dependent Gutzwiller
renormalization factors. Additionally, the chemical potential is
chosen to give the correct electron concentration, $ n=1-\delta=
\langle \opn_i \rangle_0 $. Numerical solutions of these mean fields
as a function of doping concentration are plotted in Fig.1
for realistic parameters,  $ t_x = t_y $ and $ J_x=J_y=0.3 t_x $.
Within this mean field treatment
the excitations are described by the effective
quasiparticle Hamiltonian obtained by
standard Bogolyubov transformation of $ {\cal H}_{MF} $, $ {\cal H}_{eff} =
\sum_{\vk,\sigma=1,2}  E_{\vk}  \gamma^{\dag}_{\vk \sigma} \gamma_{\vk
\sigma} $ with the spectrum
$ E^2_{\vk} = \varepsilon^2_{\vk} + \Delta^2_{\vk} $. The
quasiparticles define the the ground state by the condition $
\gamma_{\vk,\sigma} | \Phi_0 \rangle =0 $. For finite
$ \Delta_{\vk} $ these excitations have always a finite gap.

We begin at half-filling where this system reduces to a Heisenberg
ladder. This case was considered by various groups
for different couplings along the chains, $ J_x $, and  rungs $
J_y $ [1,8,9]. In the limit $ J_y \gg J_x > 0 $ the ground state consists
essentially of singlet dimers on each rung leading to a spin liquid
(singlet) ground state. The lowest spin
(triplet) excitation is obtained by replacing one singlet by
a triplet dimer with an excitation energy $ \sim J_y - J_x $.
By numerical diagonalization and dimer mean field treatments it
was demonstrated that the spin liquid ground state persists
even for $ J_x \sim J_y $ with a spin gap of $ 0.4 - 0.6 J $ for $
J=J_x=J_y $ [8,9].

The mean field solution at
half-filling independent of the coupling constants satisfies the
relation $ \Delta_x \Delta_y
+ \chi_x \chi_y = 0 $. This state is similar to the d-wave RVB in the
square lattice, where the pairing mean fields $ \Delta_{\alpha} $
differ by a phase $ \pi $ in $ x $- and $ y $-direction. It is also
identical to the Affleck-Marston flux phase [10] with
half integer flux quanta as follows from a SU(2) symmetry [6].
Upon doping, the flux phase and the d-wave RVB state differ.
In the case of the square lattice, the d-wave RVB
state is found to have lower energy than the staggered flux
phase [11]. We expect the similar result for the ladder,
and will not discuss the flux phase here.

Turning to the spin excitation (singlet-triplet) we find that it is not well
described by the simple quasiparticle Hamiltonian $ {\cal H}_{eff} $
given above. The spin gap is generally too large ($ \approx 2 J $ for
$ J_x = J_y $). Additionally the excitation energy has several minima
at the momenta $ \vq = (0,0),(0,\pi),
(\pi,0) $ and $ (\pi,\pi) $, while in the theories mentioned above
the only minimum lies at $ \vq = (\pi,\pi) $ [8,9].

We can remedy this flaw in our treatment
partially by taking the residual interaction
among quasiparticles into account. Most easily the spectrum of the spin
excitations can be obtained from the dynamical (transverse) spin
susceptibility,
$ \chi(\omega,\vq) = \langle \langle S^{-}_{-\vq}, S^{+}_{\vq} \rangle
\rangle_{\omega} $. We use the equation of motion to determine $
\chi(\omega,\vq) $ on the level of RPA including the residual
interaction. In Fig.2 we show the spectrum $ Im \chi(\omega, \vq) $
at $ \vq =(\pi, \pi) $ for
both the simple mean field and the RPA corrected result with $ J_x =
J_y $. The RPA shows a sharp excitation peak at $ \approx 0.15 J_x $
and a broad continuous spectrum between about 2$ J_x $ - 3$ J_x $ which is
the strongly suppressed remainder of the continuous excitation spectrum of $
{\cal H}_{eff} $.
The analysis of the full $ \vq $-dependence shows that the minimal
excitation is obtained at $ \vq = (\pi , \pi) $. The corresponding
binding energy is rather large bringing the excitation gap down to a
value of $ 0.15 J_x $ which is only slightly smaller than the
one found in other calculations ($ \approx 0.5 J_x $) [8,9].

As a consequence of the local constraint the singlet dimer on a rung
is a coherent superposition of pair states in the bonding ($
c^{\dag}_{i+,{\rm s}} = (c^{\dag}_{i,1,{\rm s}} +
c^{\dag}_{i,2,{\rm s}})/\sqrt{2} $) and the antibonding ($
c^{\dag}_{i-,{\rm s}} = (c^{\dag}_{i,1,{\rm s}} -
c^{\dag}_{i,2,{\rm s}})/\sqrt{2} $) state, i.e.: avoiding
double occupancy the rung state $
(1/\sqrt{2}) (c^{\dag}_{i+ \uparrow} c^{\dag}_{i+ \downarrow} -
c^{\dag}_{i- \uparrow} c^{\dag}_{i- \downarrow}) |0 \rangle $
gains the maximal exchange energy. In other words,
the splitting of the bonding and antibonding band, which would appear
in a band structure description, is absent in the present of the local
constraint so that they are equally filled at half-filling.

To get an intuitive understanding of the properties of this sharp
excitation found in RPA let us consider the problem in
the strong coupling limit, $ J_y \gg J_x $. In the mean field
a triplet excitation
creates two quasiparticles which may be considered as spinons
freely moving and destroying each one singlet dimer.
The energy loss is $ 3J_y /2 $,
neglecting the kinetic
energy contribution. These two quasiparticles can
form a triplet dimer on a rung with the lower energy $ J_y $.
Thus, there exists an effective attraction between them
leading to a bound exciton state within the quasiparticle
excitation gap. This interaction has only an attractive triplet channel if
one of the two quasiparticles is in the bonding and the
other in the antibonding band. Furthermore, the kinetic energy of the
triplet dimer is given by $ \tilde{J}_y {\rm cos} q_x $ which is lowest
at $ q_x = \pi $.
Hence, we interpret the sharp excitation in the RPA
calculation for $ J_x = J_y $ as an exciton with lowest energy
at $ \vq = (\pi , \pi) $ and the reduced continuum above $ \sim 2 J_x
$ as a spinon continuum (Fig.2).

Upon doping holes the two quasiparticle bands which are degenerate at
half-filling split. The antibonding band is raised and the bonding
band slightly lowered relative to the chemical potential.
As holes are doped the kinetic energy
can be lowered if they occupy preferentially the antibonding band.
In this way the antibonding band is gradually emptied down to a
critical doping $ \delta_c $ where eventually only the bonding band is
occupied.
The loss of antibonding electrons leads to a gradual decrease of the
pairing amplitude $ \Delta_{\alpha} $ which disappears
eventually at $ \delta_c $. The long tail of finite $ \Delta_{\alpha}
$ down to $ \delta_c $ may be an artifact of our approximation.

The behavior of $ \Delta_{\alpha} $ yields a monotonic decrease of the
quasiparticle excitation gap described by $ {\cal H}_{eff} $. On the
other hand, the spin excitation band formed by the triplet exciton
is shifting non-monotonically with doping leading to an {\it increase}
in the spin gap at small doping (see Fig.3, for coupling
constants $ t_x=t_y=t $ and $ J_x=J_y=J=0.3 t $). With the shift of the
Fermi levels in the antibonding band the optimal relative momentum
$ q_x $ for a triplet exciton of two quasiparticles deviates
gradually from $ \pi $. Since $ q_x = \pi $ yields the strongest attraction
this deviation weakens the interaction and decreases the
binding energy. The resulting increase of the spin gap for small
doping shows that here hole doping suppresses antiferromagnetic
correlation and stabilizes a spin liquid state. Also results of exact
diagonalization confirm this tendency [12].

The BCS superconductivity order parameter is given by the pairing
mean field  multiplied by the Gutzwiller renormalization factor $ g_{t
\alpha} $ as shown in Ref.6. In Fig.4 we
show the two components of $ \Delta_{SC \alpha} = \Delta_{\alpha}
g_{t \alpha} $. There is a phase difference of $
\pi $ between $ \Delta_{SC x} $ and $ \Delta_{SC y} $, reminding
of d-wave pairing, although d-wave is a misnomer in this low symmetry
system.

In summary, our mean field theory of the $ t-J
$-ladder model gives a description of the doped
spin liquid system. The spin liquid state persists for a
finite doping region away from half-filling and seems even to become
more stable with weak doping. Although mean field is
certainly not a good concept for quasi 1D system, we have seen that
qualitatively reasonable results are obtained.
We may expect that weak interchain (intraplanar
and interplanar) coupling
would lead essentially to a 3D situation
stabilizing the mean field solution.
The superconductivity obtained in this theory is intimately connected
with the existence of a spin liquid state (see also in Ref.2 and 13).
Considering their energy
scales we observe that at light doping the spin gap is by
far larger than the superconducting pair correlation energy.
Beside the triplet exciton mode discussed here we expect in the doped
region a low lying collective (sound) mode in connection with the
superconducting order will occur [14].

Our result suggests that also for chains extended in $ y $-direction
having 4, 6 or a larger even number of parallel chains would
qualitatively be very similar. Although the spin gap must be diminished
with growing system size, as recent mean field calculations show [9],
the concept of the doped spin liquid system is still applicable.
In this sense the superconducting state would continuously tend to a real
d-wave state, if we would extend the lattice in $ y $-direction
approaching in this way the 2D square
lattice.

\vskip 1.5 cm

\noindent
{\bf Acknowledgment:} We are grateful to H. Tsunetsugu, S. Gopalan, G.
Blatter, D. Khveshenko, R. Hlubina and C. Albanese for helpful and
stimulating discussions. We like also to express our gratitude to the
Swiss National Science Foundation for financial support.

\newpage

\noindent
{\bf References} \\

\noindent
\begin{description}
\item[[1]]  S.P. Strong and A.J. Millis, Phys. Rev. Lett. {\bf 69}, 2419
       (1992)
\item[[2]]  M. Ogata et al.,,Phys. Rev. {\bf B44}, 12083 (1991); M.
            Ogata et al., Phys. Rev. Lett. {\bf 66}, 2388 (1991); M.
            Imada, Phys. Rev. {\bf B48}, 550 (1993)
\item[[3]]  M. Takano, Z. Hiroi, M. Azuma, and y. Takeda, Jap. J. of Appl.
     Phys. Series {\bf 7}, 3 (1992)
\item[[4]] T.M. Rice, S. Gopalan, and M. Sigrist, Europhys.Lett. {\bf
23},445 (1993).
\item[[5]] P.W. Anderson, Science {\bf 235}, 1196 (1987)
\item[[6]]  F.C. Zhang, C. Gros, T.M. Rice, and H. Shiba, Supercond.
Sci. Technol. {\bf 1}, 36 (1988)
\item[[7]] T.C. Hsu, Phys. Rev. {\bf B41}, 11379 (1990)
\item[[8]]  E. Dagotto, J. Riera, and D. Scalapino, Phys. Rev. {\bf B45},
       5744 (1992); T. Barnes et al., Phys. Rev. {\bf B47}, 3196 (1993)
\item[[9]] S. Gopalan et al., unpublished
\item[[10]] I. Affleck and J.B. Marston, Phys. Rev. {\bf B37}, 3774
(1988)
\item[[11]] F.C. Zhang, Phys. Rev. Lett. {\bf 64}, 974 (1990); G.J.
            Chen et al., J. Phys.: Condens. Matter {\bf 3}, 5213
            (1991); M.U. Ubben and P.A. Lee, Phys. Rev. {\bf B46},
            8434 (1992)
\item[[12]]  E. Dagotto et al., Phys. Rev. {\bf B45},10741 (1992)
\item[[13]] J. Fr\"ohlich and U. Studer, Rev. Mod. Phys. {\bf 65}, 733
(1993)
\item[[14]] M. Sigrist et al., unpublished
\end{description}

\vskip 2 cm

\noindent
{\bf Figures Caption} \\

\vskip 0.7 cm

\noindent
{\bf Fig.1:} The mean fields as a function of the doping
concentration for $ J_x=J_y=J=0.3 t $.

\vskip 1 cm

\noindent
{\bf Fig.2:} The spectrum of the spin-excitations for $ J_x=J_y=J=0.3
t $ and $ \vq = (\pi,\pi) $. The solid line
denotes the exciton spectrum obtained from RPA and the dashed line the
mean field quasiparticle spectrum neglecting the interaction among the
quasiparticles. The sharp exciton mode at $ \omega \approx 0.15 J $ contains
practically all weight of the excitation spectrum.

\vskip 1 cm

\noindent
{\bf Fig.3:} The spin gap as a function of the doping concentration
for$ J_x=J_y=J=0.3 t $.
The solid line denotes the bottom of the quasiparticle spectrum and
the dashed line the gap of the exciton state.

\vskip 1 cm

\noindent
{\bf Fig.4:} The superconductivity order parameter versus the doping
concentration for $ J_x=J_y=J=0.3 t $.

\end{document}